\documentclass[sigconf]{acmart}
\usepackage{balance}

\copyrightyear{2026}
\acmYear{2026}

\setcopyright{cc}
\setcctype{by}

\copyrightyear{2026}
\acmYear{2026}

\setcopyright{cc}
\setcctype{by}

\acmConference[UbiComp Companion '26]
{Companion of the 2026 ACM International Joint Conference on Pervasive and Ubiquitous Computing}
{October 11--15, 2026}
{Shanghai, China.}

\acmBooktitle{Companion of the 2026 ACM International Joint Conference on Pervasive and Ubiquitous Computing (UbiComp Companion '26), October 11--15, 2026, Shanghai, China}

\acmISBN{979-8-4007-2539-5/2026/10}
\acmDOI{10.1145/3798063.3837270}

\graphicspath{{./}}

\newcommand{\missingfigure}[2]{%
  \fbox{%
    \parbox[c][#1][c]{#2}{%
      \centering
      Figure file not found.\\
      Upload the corresponding PDF to the project.
    }%
  }%
}

\AtBeginDocument{%
  }

\begin{document}

\title[ECHOES]{ECHOES: An Adaptive LLM-Powered AR Glasses Docent for Evidence-Grounded Urban Heritage Interpretation}

\author{Bang Xiao}
\affiliation{%
  \institution{Digital Heritage Lab,}
  \institution{Center for Data Science}
  \institution{New York University Shanghai}
  \city{Shanghai}
  \country{China}}
\email{bx2181@nyu.edu}

\author{Baosheng Jin}
\affiliation{%
  \institution{Digital Heritage Lab,}
  \institution{Center for Data Science}
  \institution{New York University Shanghai}
  \city{Shanghai}
  \country{China}}
\email{bj2410@nyu.edu}

\author{Jinhan Niu}
\affiliation{%
  \institution{Digital Heritage Lab,}
  \institution{Center for Data Science}
  \institution{New York University Shanghai}
  \city{Shanghai}
  \country{China}}
\email{jn3063@nyu.edu}

\author{Xingchen Zhang}
\affiliation{%
  \institution{Digital Heritage Lab,}
  \institution{Interactive Media Arts}
  \institution{New York University Shanghai}
  \city{Shanghai}
  \country{China}}
\email{xz33@nyu.edu}

\author{Hua Shen}
\affiliation{%
  \institution{Center for Data Science}
  \institution{New York University Shanghai}
  \city{Shanghai}
  \country{China}}
\email{huashen@nyu.edu}

\renewcommand{\shortauthors}{Bang Xiao, Baosheng Jin, Jinhan Niu, Xingchen Zhang, and Hua Shen}

\begin{abstract}
Urban heritage tours are often delivered through fixed scripts, leaving limited room for visitors' situated questions, language preferences, or requests for evidence. We present ECHOES, an \emph{adaptive} LLM-powered AR glasses docent for evidence-grounded interpretation of historical buildings.
Specifically, the agent \emph{adapts} along three dimensions: it modifies response structure according to the type of visitor question (identification, architectural detail, evidence-seeking, or route inquiry); it accumulates session-level interest signals from repeated questions and expressed curiosity to shape later explanations; and it tracks language preference and retrieval context across turns to maintain conversational continuity. The core of ECHOES is an LLM agent initialized as a source-aware urban heritage docent that realizes this adaptation through dialogue and preference context maintained across the session. Around this agent, ECHOES integrates visual site grounding, a self-built historical knowledge base supporting retrieval-augmented generation, and concise multilingual AR output designed for glanceable use. Through a working prototype centered on Shanghai's East Nanjing Road and the Bund, we demonstrate how an adaptive LLM-powered AR docent can transform historical tours from scripted narration into situated, question-responsive, and source-aware conversations.
\end{abstract}



\maketitle

\begin{figure*}[t]
  \centering
  \IfFileExists{figure1.pdf}{%
    \includegraphics[width=\textwidth]{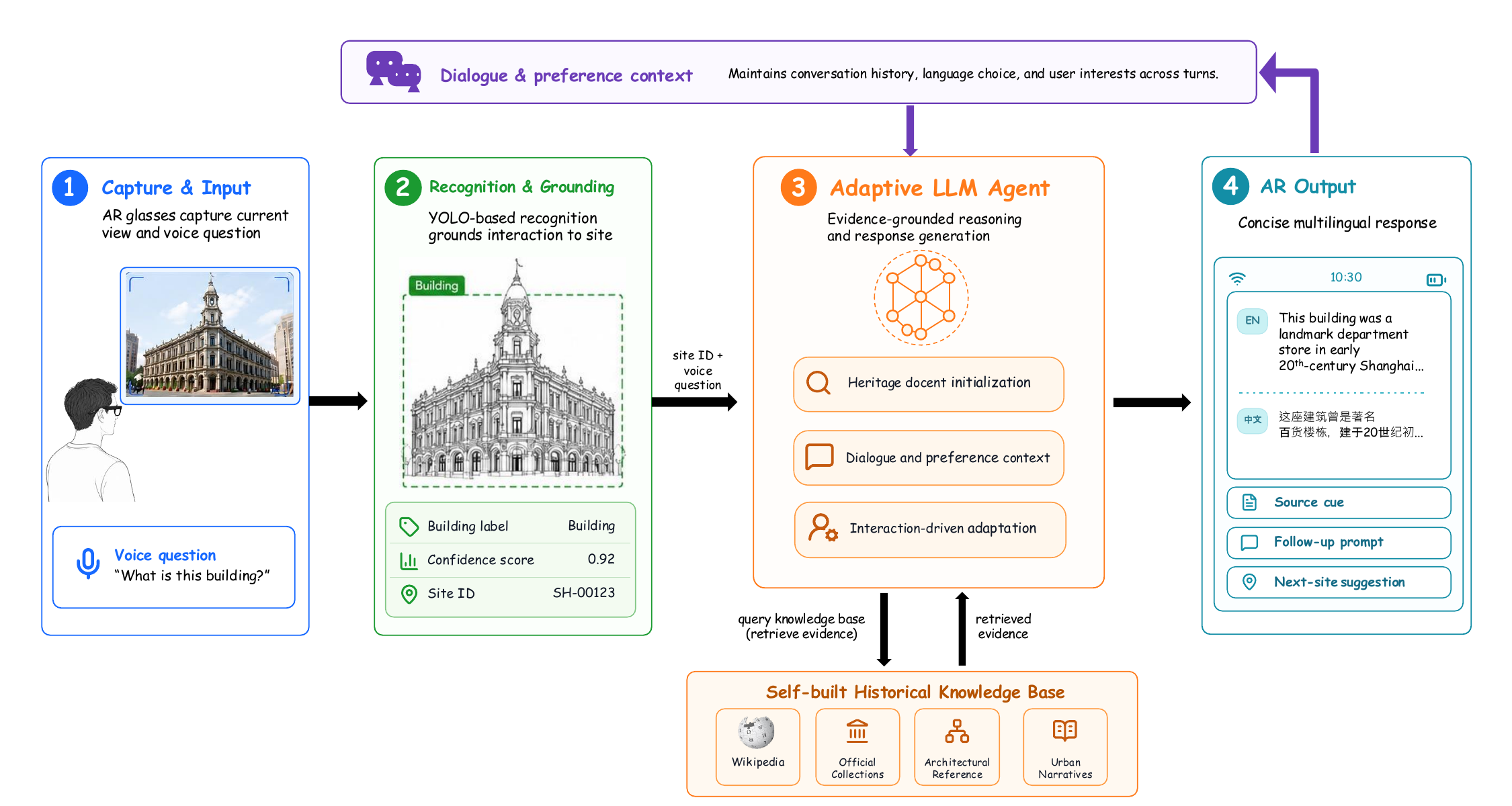}%
  }{%
    \missingfigure{1.8in}{0.95\textwidth}%
  }
  \caption{System overview of ECHOES. Capture and input, recognition and grounding, the historical knowledge base, and AR output are organized around the adaptive LLM agent, which maintains dialogue and preference context across turns.}
  \Description{A wide system overview figure showing the ECHOES architecture. Supporting components include capture and voice input, recognition and site grounding, a self-built historical knowledge base, and AR output. The central component is the adaptive LLM agent, which uses session context, language preference, user interests, and retrieved evidence to generate responses.}
  \label{fig:system}
\end{figure*}

\section{Introduction}

Historical tours are commonly delivered as scripted narration. In museums, heritage sites, and historical streets, visitors receive a predetermined sequence of descriptions through guidebooks, audio guides, or mobile applications. This format efficiently communicates curated content but leaves little room for the situated questions that emerge from what visitors actually notice on site. A visitor standing in front of a historical building may ask not only ``what is this building?'' but also ``why was it famous?'', ``can you explain the facade?'', ``is this interpretation evidence-based?'', or ``what related place should I visit next?'' These questions vary in type — identification, architectural detail, evidence-seeking, route inquiry — and \textbf{shift unpredictably across a visit}. Scripted systems cannot flexibly respond to such variation in curiosity, language preference, and evidence-seeking behavior.


AR and mixed reality technologies have been widely explored for cultural heritage, including immersive reconstructions, location-based storytelling, and object-centered interpretation \cite{ramtohul2024ar,trichopoulos2024smartglasses,litvak2020enhancing,dima2022design}. AR glasses are particularly promising for historical building tours because they allow visitors to keep their attention on the surrounding site while receiving lightweight contextual information. Meanwhile, large language models (LLMs) further enable natural language dialogue, multilingual interaction, and flexible content generation in tour-guidance and wearable-assistant contexts \cite{wang2024virtuwander,lee2024gazepointar,janaka2024tom,cai2025aiget}. 
However, connecting an LLM to an AR interface alone is insufficient for historical interpretation: a heritage docent must ground open-ended dialogue in specific sites and source materials, maintain the visitor's expressed interests and language preference across turns, and adapt its explanations to the type of question being asked — not merely generate fluent text on demand.


We present ECHOES, an adaptive LLM-powered AR glasses docent for evidence-grounded urban heritage interpretation. 
The central idea is to replace fixed-script narration with an agent loop that adapts to the visitor \textbf{in three ways}: it adjusts response structure to the \textbf{type of question} asked (identification, architectural detail, evidence-seeking, or route inquiry); it accumulates \textbf{session-level interest signals} from repeated questions and expressed curiosity to shape later explanations; and it tracks \textbf{language preference and retrieval context} across turns to maintain conversational continuity. Rather than delivering a predetermined tour, ECHOES lets visitors follow their own curiosity — in whatever language they prefer — while responses remain grounded in historical sources and concise enough for AR display.

%


This work makes four contributions: (1) an adaptive LLM agent that coordinates visitor question type, session-level interests, language preference, and retrieval context across turns; (2) a self-built historical knowledge base supporting retrieval-augmented, source-traceable heritage interpretation; (3) an interaction-type taxonomy mapping visitor question types to distinct response structures; and (4) a working end-to-end prototype integrating visual site grounding, RAG, and an AR output schema designed for the constraints of glasses displays.


\section{ECHOES System}

ECHOES consists of three supporting components organized around the adaptive LLM agent: site grounding and voice activation, a self-built historical knowledge base, and multilingual AR response generation. Figure~\ref{fig:system} illustrates the system architecture. These components address the practical problems users face during situated heritage tours: they may not know the name of the building in front of them, they may want flexible answers without losing historical grounding, and they may need short multilingual guidance without looking down at a phone.

\subsection{Site Grounding and Voice Activation}

The first component grounds the conversation in the historical site being viewed. In the current prototype, AR glasses capture the user's situated view and voice question. A YOLO-based recognition module maps the viewed building or selected image to a supported site ID, building label, and confidence score \cite{redmon2016you}. The system currently supports a curated set of historical buildings along Shanghai's East Nanjing Road and the Bund area, including the Wing On Company Building, Peace Hotel, Shanghai No.1 Department Store, First Food Store, Shanghai Fashion Store, and other commercial landmarks. This component addresses the user problem of not knowing how to name or search for the building in front of them: instead of manually looking up a location, the visitor can look at a building and ask a question. Once the site is grounded, follow-up questions such as ``why was it famous?'' or ``what happened here?'' are interpreted relative to the current building context.

\subsection{Self-Built Historical Knowledge Base}

The second component provides the evidence substrate for open-ended historical interpretation. ECHOES is grounded in a self-built historical knowledge base for supported Shanghai historical buildings. The knowledge base integrates heterogeneous materials, including Wikipedia entries, official East Nanjing Road historical-building collections, architectural references such as \textit{Shanghai Famous Buildings} / \textit{Shanghai Ming Jianzhu Zhi}, and selected online narratives about urban memory, commercial life, and local stories. Each source item is organized with metadata such as building name, alternative names, source type, language, time period, theme tags, and source excerpt.

This component addresses the user problem of wanting flexible explanations without receiving unsupported historical claims. ECHOES uses retrieval-augmented generation (RAG) to connect user questions with relevant source materials \cite{lewis2020rag,asai2024selfrag,huang2026survey}. Official and architectural references are used to support factual claims, encyclopedic sources provide background information and multilingual names, and online urban narratives enrich the explanation with local memory and everyday-life stories. When users ask evidence-seeking questions such as ``is there evidence for that?'', ECHOES can surface source cues indicating whether the response draws on historical building records, architectural references, encyclopedic materials, or selected urban narratives.

\subsection{Multilingual AR Response}

The third component formats agent outputs for AR use. ECHOES supports multilingual input and output: visitors can ask questions in Chinese, English, or other supported languages, and the agent maintains the selected language across turns unless the user requests a switch. Prior work on multilingual retrieval-augmented systems suggests that language preference can influence retrieval and generation behavior, motivating explicit attention to multilingual interaction in grounded QA systems \cite{park2025investigating}. Because AR glasses have limited display space, responses are designed to be concise, glanceable, and easy to continue through follow-up. The AR output may include a short answer, a source cue, a follow-up prompt, or a next-site suggestion. This component addresses the user problem of receiving useful guidance without reading long text on a small display or breaking attention from the surrounding site.

\section{Adaptive LLM Agent}

The adaptive LLM agent is the central component that coordinates site context, user input, dialogue history, and AR response generation. After a building is grounded to a site ID, the agent receives the transcribed voice question, the current session context, the user's language preference, and the retrieved source materials. It then generates a response that is concise enough for AR display while remaining connected to the current site. This design allows ECHOES to support historical interpretation as a continuing interaction, where later responses can build on the user's earlier questions and expressed interests.

\textbf{Heritage docent initialization.} The agent is initialized with the role of an urban heritage docent. The prompt defines the agent as a guide for historical buildings, with instructions to answer around the current site, provide concise explanations, avoid unsupported historical claims, and respond in the user's preferred language. This initialization addresses a common problem in LLM-based guidance: a general-purpose model may produce fluent explanations, but the answer may be too broad, too long, or insufficiently connected to the building in view. In ECHOES, the docent role constrains the response style and makes the agent behave as a situated guide for the surrounding urban environment.

\textbf{Dialogue and preference context.} ECHOES maintains a session context across multiple turns. The context records the current site, previous user questions, language choice, retrieved evidence, and user-expressed interests. For example, if the visitor repeatedly asks about facades, commercial life, or everyday urban experience, these signals are stored as part of the ongoing session. If the visitor asks for evidence, the system also records this evidence-seeking behavior and can surface source cues more explicitly in later responses. This context allows the user to continue the tour without restating the same preference in every turn. It also allows the agent to maintain continuity when the conversation moves from one building to another.

\textbf{Interaction-driven adaptation.} The agent adapts its docent behavior according to the type of interaction initiated by the visitor. A direct identification question triggers a concise building introduction; a significance-oriented question triggers a historical explanation of the building's commercial, architectural, or social role; a visual-detail question triggers a facade- or street-level interpretation; an evidence-seeking question triggers a source-aware response; and a route-oriented question triggers a nearby-site recommendation. These interaction types lead to different response structures, not merely different wording. Depending on the user's action, the agent may answer directly, ask a follow-up question, show a source cue, or suggest the next place to visit. In this way, ECHOES turns open-ended user interaction into adaptive docent behavior instead of replaying a fixed sequence of descriptions.

\section{Demonstration}

\begin{figure*}[!t]
  \centering
  \IfFileExists{figure2.pdf}{%
    \includegraphics[width=\textwidth]{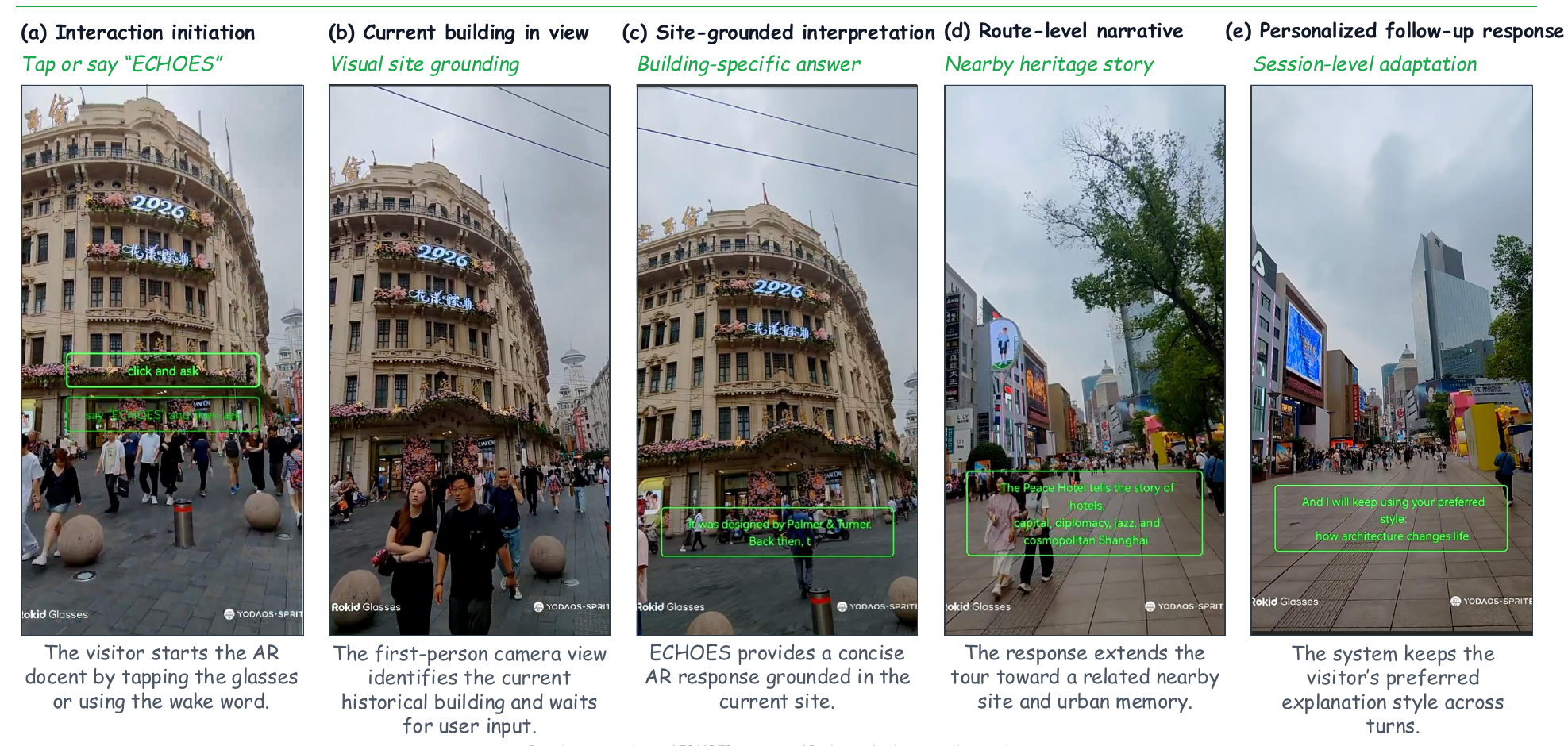}%
  }{%
    \missingfigure{1.8in}{0.95\textwidth}%
  }
  \caption{Demonstration walkthrough of ECHOES. (a) The visitor initiates the AR docent by tapping the glasses or using the wake word. (b) ECHOES grounds the current building in the first-person camera view. (c) The agent generates a concise site-specific response. (d) The response extends toward a related nearby heritage site. (e) The agent preserves the visitor's preferred explanation style across turns.}
  \Description{A five-panel demonstration figure showing ECHOES in use. The panels show interaction initiation, the current building in the first-person view, a site-grounded response, a route-level narrative toward a nearby site, and a personalized follow-up response that preserves the visitor's preferred explanation style.}
  \label{fig:demo}
\end{figure*}

We demonstrate ECHOES through an AR-glasses prototype with a companion display that mirrors the visitor's first-person interaction. The demonstration centers on historical buildings along Shanghai's East Nanjing Road and the Bund, showing how the system initiates an AR docent session, grounds the current building, generates a site-specific response, extends the interpretation to nearby sites, and adapts follow-up responses based on the visitor's expressed preferences.

As shown in Figure~\ref{fig:demo}, the visitor starts the AR docent by tapping the glasses or using the wake word ``ECHOES.'' The system then uses the first-person camera view to identify the current building and wait for the visitor's voice question. After receiving the question, ECHOES generates a concise AR response grounded in the current site. The response can also extend the tour toward a nearby heritage site, such as the Peace Hotel, and later turns can preserve the visitor's preferred explanation style.

For the Wing On Company Building scenario, the visitor first activates the docent while facing the building. ECHOES identifies the current site and supports natural follow-up questions such as ``What is this building?'', ``Why was it famous?'', or ``Can you explain the facade?'' The agent retrieves relevant historical materials and generates a concise building-specific answer for AR display. When the visitor asks for a related place, ECHOES can extend the interpretation toward the East Nanjing Road--Bund axis and recommend the Peace Hotel as a nearby site connected through hotel culture, diplomacy, jazz, and cosmopolitan memory.

The final part of the demonstration shows session-level adaptation. If the visitor expresses a preference for explanations that connect architecture with everyday urban life, ECHOES records this preference in the dialogue context and applies it in later turns. The companion display mirrors the AR-glasses interaction so that attendees can observe the current site, generated answer, source-aware state, and follow-up prompt. Through this walkthrough,\linebreak[4]
ECHOES demonstrates how an LLM-powered AR docent can support
site-grounded questioning, evidence-aware explanation, route-level
storytelling, and personalized follow-up within a single urban
heritage tour.

\section{Conclusion and Future Work}

This paper presents ECHOES, an adaptive LLM-powered AR glasses docent for evidence-grounded urban heritage interpretation. The central contribution is the adaptive agent that coordinates site grounding, user questions, dialogue history, language preference, retrieved evidence, and AR-suitable response generation. Compared with fixed-script tours, ECHOES supports situated site grounding, open-ended voice questioning, session-level preference learning, multilingual AR response, evidence checking, and route-level recommendation.

The current prototype focuses on a curated set of Shanghai historical buildings around East Nanjing Road and the Bund. Future work will expand the historical building corpus, improve long-term preference memory beyond a single session, refine how source cues and uncertainty are presented on AR glasses, and deploy the system in real urban heritage routes. We also plan to study how adaptive LLM docents influence visitor agency, trust calibration, attention to the built environment, and historical understanding in situated cultural heritage experiences.

\begin{acks}
This work was supported by the Shanghai Pujiang Talents Program
(Grant No. 25PJA109). We also gratefully acknowledge the support
of the Center for Data Science and Digital Heritage Lab at
New York University Shanghai.
\end{acks}

\bibliographystyle{ACM-Reference-Format}
\balance
\bibliography{references}

\end{document}